\documentclass{aa}
\usepackage{multirow}
\usepackage{amsmath}
\usepackage[colorlinks,linkcolor=blue,anchorcolor=blue,citecolor=blue]{hyperref}
\usepackage{graphicx} 
\usepackage{lscape}
\usepackage{longtable}
\usepackage{txfonts}
\usepackage{natbib}
\usepackage{color}

\begin{document}

\title{X-ray analysis of JWST's first galaxy cluster lens SMACS~J0723.3-7327}

\author{
A.~Liu\inst{}\thanks{e-mail: \href{mailto:liuang@mpe.mpg.de}{\tt liuang@mpe.mpg.de}},
E.~Bulbul\inst{},
M.~E.~Ramos-Ceja\inst{}, 
J.~S.~Sanders\inst{},
V.~Ghirardini\inst{},
Y.~E.~Bahar\inst{}, 
M.~Yeung\inst{}, 
E.~Gatuzz\inst{}, 
M.~Freyberg\inst{}, 
C.~Garrel\inst{}, 
X.~Zhang\inst{}, 
A.~Merloni\inst{}, 
K.~Nandra\inst{}}

\institute{
\inst{}{Max Planck Institute for Extraterrestrial Physics, Giessenbachstrasse 1, 85748 Garching, Germany}\\
}

\titlerunning{X-ray analysis of SMACS~J0723.3-7327}
\authorrunning{Liu et al.}

\abstract
{SMACS~J0723.3-7327 is the first galaxy cluster lens observed by {\sl JWST}. Based on the Early Release Observation data from {\sl JWST}, several groups have reported the results on strong lensing analysis and mass distribution of this cluster. The new lens model dramatically improves upon previous results, thanks to the unprecedented sensitivity and angular resolution of {\sl JWST}. However, limited by the angular coverage of the {\sl JWST} data, the strong lensing models only cover the central region. X-ray analysis on the hot ICM is necessary to obtain a more complete constraint on the mass distribution in this very massive cluster.}
{In this work, we aim to perform a comprehensive X-ray analysis of J0723 to obtain accurate ICM hydrostatic mass measurements, using the X-ray data from {\sl SRG (Spectrum Roentgen Gamma)/eROSITA} and {\sl Chandra} X-ray observatories. By comparing the hydrostatic mass profile with the strong lensing model, we aim to provide the most reliable constraint on the distribution of mass up to $R_{500}$. }
{Thanks to the {\sl eROSITA} all-sky survey and {\sl Chandra} data, which provide high S/N and high angular resolution respectively, we are able to constrain the ICM gas density profile and temperature profile with good accuracy both in the core and to the outskirts. With the density and temperature profiles, we computed the hydrostatic mass profile, which was then projected along the line of sight to compare with the mass distribution obtained from the recent strong lensing analysis based on {\sl JWST} data. We also deprojected the strong lensing mass distribution using the hydrostatic mass profile we obtained in this work. }
{The X-ray results obtained from {\sl eROSITA} and {\sl Chandra} agree very well with each other. The hydrostatic mass profiles we measured in this work, both projected and deprojected, are in good agreement with recent strong lensing results based on {\sl JWST} data, at all radii. The projected hydrostatic mass within 128~kpc (the estimated Einstein radius) is $(8.0\pm0.7) \times 10^{13} M_{\odot}$, consistent with the strong lensing mass reported in \citet{2022Caminha}: $(8.6\pm0.2) \times 10^{13} M_{\odot}$, and in \citet{2022Mahler}: $(7.6\pm0.2) \times 10^{13} M_{\odot}$. With the hydrostatic mass profile, we measure $R_{2500}=0.54\pm0.04$~Mpc and $M_{2500}=(3.5\pm0.8) \times 10^{14}~M_{\odot}$, while the $R_{500}$ and $M_{500}$ are $1.32\pm0.23$~Mpc and $(9.8\pm5.1) \times 10^{14}~M_{\odot}$, with relatively larger error bar due to the rapidly decreasing S/N in the outskirts. We also find that the radial acceleration relation in J0723 is inconsistent with the RAR for spiral galaxies, implying that the latter is not a universal property of gravity across all mass scales. }
{}


\maketitle

\section{Introduction}
Mass and mass distribution in massive galaxy clusters are crucial to study the property of dark matter halos and the evolution of the large-scale structures in the Universe. 
There are several independent ways to measure the mass distribution of clusters, such as strong/weak gravitational lensing effect of background sources \citep[see, e.g.,][]{2011Kneib, 2020Umetsu, 2022Chiu, 2017Caminha, 2022Bergamini}, velocity dispersion of member galaxies \citep[e.g.,][]{1993Girardi}, hydrostatic equilibrium of the X-ray emitting intracluster medium (ICM) \citep[e.g.,][]{2013Ettori, 2018Ghirardini, 2019Ettori}, and CMB lensing \citep{2015Melin}. Among them, the X-ray hydrostatic mass has the unique advantage of acquiring the deprojected mass distribution, while the other methods can only measure the line-of-sight integrated mass. Another difference in these methods is that they are sensitive in different radial ranges. For example, strong lensing has the most prominent effect near the critical line (close to the core region, usually smaller than 0.3$R_{500}$), where ICM hydrostatic mass may suffer from bias due to non-gravitational effects such as radiative cooling and AGN feedback particularly when a strong cool core exists, and non-thermal pressure due to gas turbulence or bulk motions. Therefore, it is important to have the mass distribution in galaxy clusters measured from multiple independent methods, to understand the biases in each of them.  

\begin{figure*}
\begin{center}
\includegraphics[width=0.49\textwidth, trim=0 50 60 80, clip]{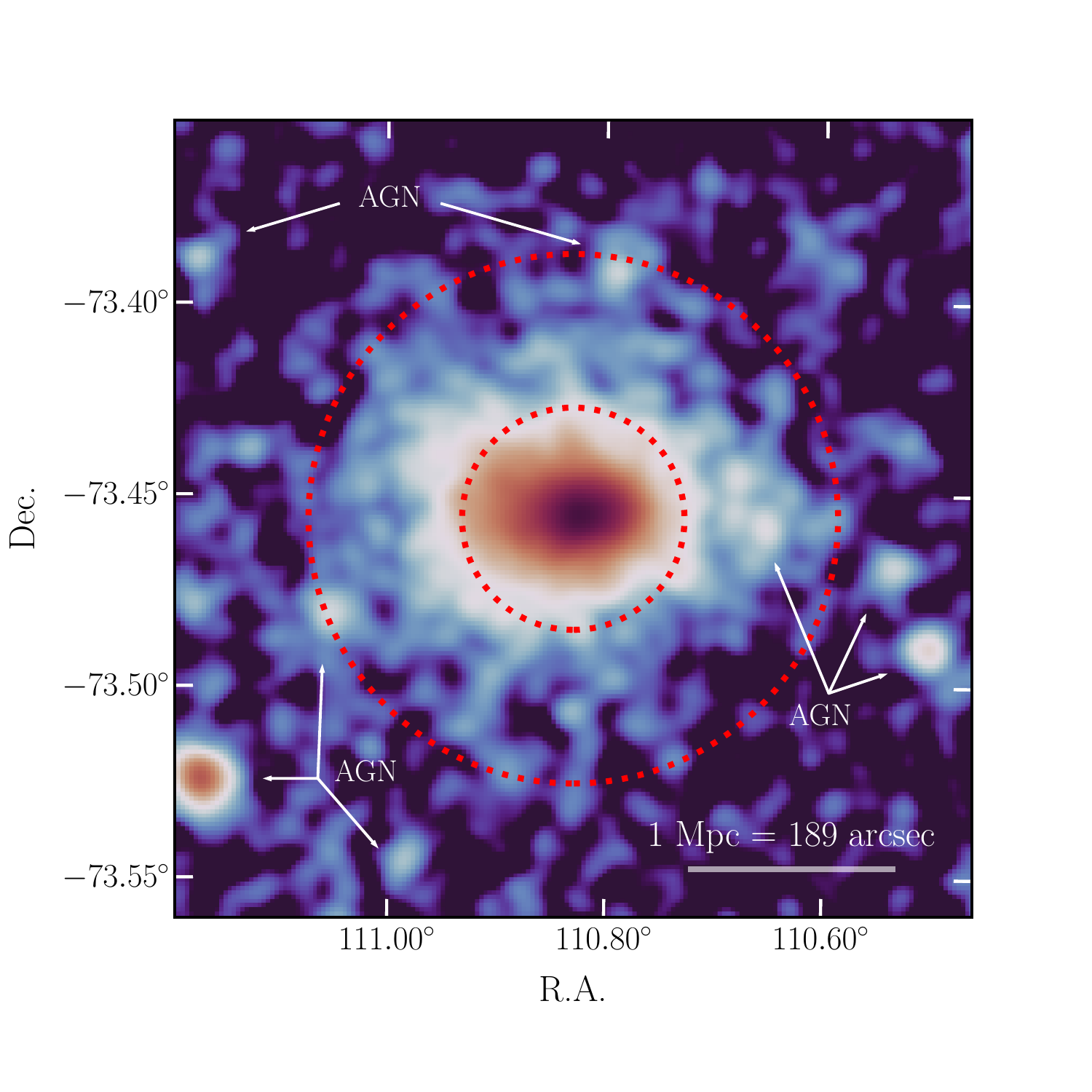}
\includegraphics[width=0.49\textwidth, trim=0 50 60 80, clip]{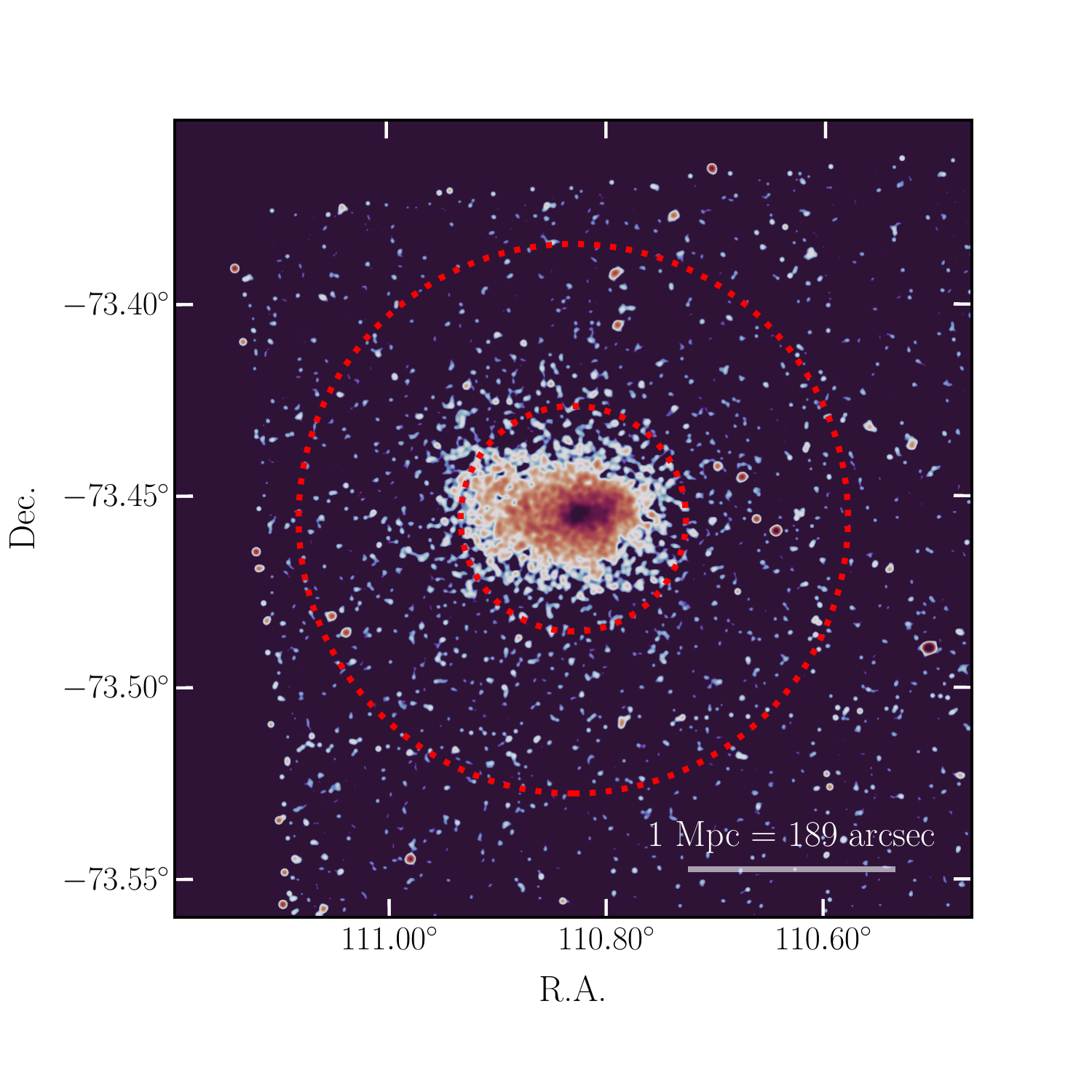}
\caption{X-ray image of J0723. {\sl Left panel:} {\sl eROSITA} exposure-corrected image in the 0.2--2.3~keV band, smoothed with a Gaussian of $\sigma=10\arcsec$. {\sl Right panel:} {\sl Chandra} exposure-corrected image in the 0.5--7~keV band, smoothed with a Gaussian of $\sigma=2\arcsec$. The images are centered at the coordinate (R.A. = 7:23:19.20, Dec. = $-$73:27:22.5). The large and small red circles show the positions of $R_{500}$=1.32~Mpc and $R_{2500}$=0.54~Mpc, determined from the hydrostatic mass profile ($R_{500}/R_{2500}$ is the radius within which the average density is 500/2500 times the critical density at cluster’s redshift, see Sect.~\ref{sec:analysis}). On the {\sl eROSITA} image, the emission of the cluster clearly extends to $\sim R_{500}$. On the {\sl Chandra} image, the emission only slightly exceeds $R_{2500}$, but the cluster core is better resolved. The red point sources on the {\sl Chandra} image (visible as larger white sources on the {\sl eROSITA} image due to the larger PSF) are AGN which are masked in the analysis. Some bright point sources are marked on the {\sl eROSITA} image. }
\label{image}
\end{center}
\end{figure*}

In this work, we will focus on a specific target, SMACS~J0723.3-7327 (hereafter J0723), a massive galaxy cluster at redshift 0.39 \citep{2001Ebeling}. J0723 (R.A. = 7:23:19.20, Dec. = $-$73:27:22.5) was identified as a galaxy cluster in the southern extension of the Massive Cluster Survey \citep[MACS;][]{2001Ebeling}. It was also detected in X-ray band in the {\sl ROSAT} All-Sky Survey \citep[RASS;][]{1999voges}, and in Sunyaev-Zeldovich survey by {\sl Planck} \citep{planck2014,2016planckb}.
J0723 is the first galaxy cluster lens observed by the James Webb Space Telescope ({\sl JWST}). Based on the Early Release Observation data from {\sl JWST}, several groups have reported the results on strong lensing analysis and mass distribution of this cluster \citep{2022Mahler, 2022Caminha}. The new lens model dramatically improves upon previous results, thanks to the unprecedented sensitivity and angular resolution of {\sl JWST}, which lead to the discovery of significantly more multiple images and lensed sources. However, the {\sl JWST} observation only covers the core of the cluster, which limits the strong lensing analysis to the central $\sim200$~kpc \citep{2022Caminha}, despite that \citet{2022Mahler} extrapolated the model to $\sim1$~Mpc. Since the strong lensing effect is only sensitive to the projected mass, the small coverage of the strong lensing model implies that, with the strong lensing results only, the mass distribution in cluster outskirts is unknown, hence one cannot get the true/deprojected mass distribution in cluster center. Therefore, to obtain a more complete constraint on the mass distribution in J0723, the strong lensing model needs to be combined with other probes such as ICM hydrostatic analysis, which provides constraints on the mass distribution from the center to the outskirts. 

J0723 has been observed in the X-ray band by {\sl Chandra} and {\sl XMM-Newton}. Using the archival data of {\sl XMM-Newton}, \citet{2020Lovisari} measured the hydrostatic mass of J0723, but with only three bins in the temperature profile. Their results show that J0723 is massive ($M_{500}=10.10_{-1.23}^{+1.57}\times 10^{14} M_{\odot}$) and hot ($kT_{500}=7.53_{-0.53}^{+0.53}$~keV). J0723 is also observed by {\sl eROSITA} during its survey observation phase.
In this work, we will perform an X-ray study on J0723 based on the {\sl eROSITA} all-sky survey data, aiming at providing the projected and deprojected hydrostatic mass profiles, to the largest extent allowed by the data.
Due to the relatively low angular resolution of {\sl eROSITA} (FOV averaged half energy width HEW$\sim$30\arcsec--35\arcsec over the soft band), we will also leverage the high angular resolution of the archival {\sl Chandra} data to constrain the temperature profile in cluster center. On the other hand, the archival {\sl Chandra} data is too shallow to constrain gas properties in cluster outskirts (see Fig.~\ref{image}). The {\sl eROSITA} all-sky survey data, thanks to its large effective area in soft energies, is most sensitive to low temperature gas at cluster outskirts ($\sim$1400~cm$^2$ at 1~keV). Therefore, it will play a crucial role to determine the gas density and temperature profiles up to $\sim R_{500}$, which are key information in the conversion between projected and deprojected mass profiles. The results in this work will also be helpful to further verify the performance of {\sl eROSITA}, improving our understanding of the calibration systematics in the analysis of {\sl eROSITA} survey-mode data on very hot clusters ($\sim$10~keV for J0723).

The paper is organized as follows. In Sect.~\ref{sec:reduction}, we introduce the {\sl eROSITA} and {\sl Chandra} X-ray datasets we used in this work and the reduction of these data. In Sect.~\ref{sec:analysis}, we present the X-ray data analysis, including the spectral and image fitting, and the computation of the mass profile. In Sect.~\ref{sec:discussion}, we discuss our results and possible caveats in our analysis. Our conclusions are summarized in Sect.~\ref{sec:conclusions}. Throughout this paper, we adopt the concordance $\Lambda$CDM cosmology with $\Omega_{\Lambda} =0.7$, $\Omega_{\mathrm m} =0.3$, and $H_0 = 70$~km~s$^{-1}$~Mpc$^{-1}$. At redshift $z=0.39$, 1~arcsec corresponds to 5.29~kpc. The solar abundance table from \citet{asplund2009} is adopted. X-ray spectral fitting in this work is done with {\tt Xspec} version 12.11.1 \citep{1996Arnaud} using C-statistics \citep{cash1979,1984Baker,2017Kaastra}. Quoted error bars correspond to a 1$\sigma$ confidence level unless noted otherwise.

\section{X-ray data reduction}
\label{sec:reduction}

In this work, we used the X-ray data from {\sl eROSITA} and {\sl Chandra}. The {\sl eROSITA} data are obtained during the all-sky survey (eRASS) from December 2019 to February 2022. J0723 was scanned five times in the first five all-sky surveys (eRASS:5). The total effective exposure after correcting for vignetting effect is $\sim4.4$~ks. We note that the exposure on J0723 is significantly larger than the average depth of eRASS, because it lies close to the south ecliptic pole of the survey. The {\sl Chandra} observation (ObsID: 15296, PI: S. Murray) was performed in April 2014, with a cleaned exposure time of 16.5~ks. 

\subsection{eROSITA}
The eRASS:5 data are processed with the eROSITA Science Analysis Software System \citep[eSASS,][]{2022Brunner}\footnote{version {\tt eSASSusers\_211214\_0\_4}.}. For all the seven telescope modules (TMs), we apply pattern recognition and energy calibration to produce calibrated event lists. In comparison with the data processing c001 from the eROSITA Early Data Release\footnote{https://erosita.mpe.mpg.de/edr/}, the data processing we performed in this work has an improved low energy detector noise suppression, and a better computation of the subpixel position. The event lists are further filtered after the determination of good time intervals, dead times, corrupted events and frames, and bad pixels. Using star-tracker and gyro data, celestial coordinates are assigned to the reconstructed X-ray photons, which can then be projected into the sky to produce images and exposure maps. All valid pixel patterns are selected in this work, i.e., single, double, triple, and quadruple events. The vignetting and point spread function (PSF) calibrations are less accurate in the corners of CCDs, therefore we use only photons which are detected at off-axis angles $\leq$ 30~arcmin. The soft band of TMs 5 and 7 are affected by light leak \citep{2021Predehl}, therefore we ignored the events below 1~keV of TMs 5 and 7 in our analysis. Point sources are detected in the {\sl eROSITA} soft band (0.2--2.3~keV) image using the same method described in \citet{2022Brunner}, and are masked from the analysis. We also include the point source list as detected by {\sl Chandra}. The radius of the masked point source has the minimum value of 30$\arcsec$. For several bright point sources, the radius was increased manually to mask the residual emission due to the outer wings of the PSF. We checked the masked regions visually and confirmed that after masking, the residual contamination from point sources is negligible.

\subsection{Chandra}
The reduction of {\sl Chandra} data is performed using the software {\tt CIAO v4.14}, with the latest release of the {\sl Chandra} Calibration Database at the time of writing (CALDB v4.9). Time intervals with a high background level are filtered out by performing a 2$\sigma$ clipping of the
light curve in the 2.3--7.3~keV band source-free image. The total cleaned exposure time after deflare is $\sim$16.5~ks. 

Point sources within the field of view (FOV) were identified with {\tt wavdetect} on the {\sl Chandra} 0.5--7~keV band image, checked visually, and eventually masked. Gaps between the CCD chips are also masked. 

The {\sl eROSITA} and {\sl Chandra} images of J0723 are shown in Fig.~\ref{image}. We note that, while the ICM emission in cluster outskirts can be well detected by the {\sl eROSITA} data, it is beyond the flux limit of the {\sl Chandra} data, despite that the {\sl Chandra} data will also play an important role in constraining the ICM properties in the core. Therefore, in the following analysis, we will consider the {\sl Chandra} results in the outskirts ($>\sim$600~kpc) as an extrapolation, and rely on the {\sl eROSITA} measurements in these regions.

\section{X-ray data analysis}
\label{sec:analysis}
In this section, we will introduce our X-ray analysis for both {\sl eROSITA} and {\sl Chandra}. We perform spectral analysis to measure the gas temperature profile, and imaging analysis to measure the gas density. The mass profile is then derived from the temperature and density profiles using the hydrostatic equilibrium equation. 

\begin{figure}
\begin{center}
\includegraphics[width=0.49\textwidth, trim=40 45 60 30, clip]{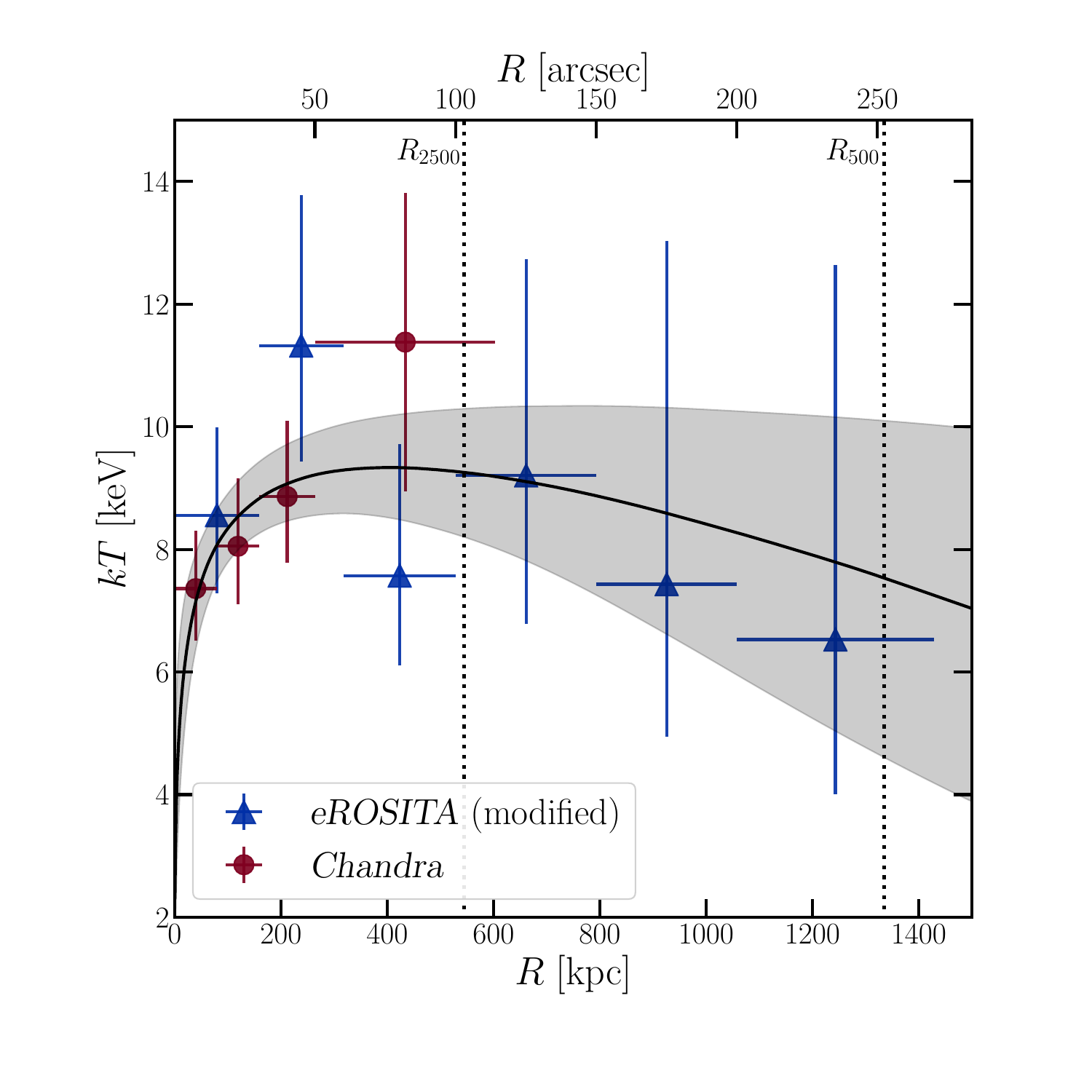}
\caption{Temperature profile of J0723. Blue and red data points are from the analysis of {\sl eROSITA} (after  modification using hard-band results) and {\sl Chandra}, respectively. The solid black curve and shaded area are the best-fit model of the profile. The two dotted lines indicate the positions of $R_{500}$=1.32~Mpc and $R_{2500}$=0.54~Mpc, determined from the hydrostatic mass profile measured by {\sl eROSITA} (the blue line in Fig.~\ref{hydromass}, see Sect.~\ref{sec:hydro}, same for the other figures). }
\label{tprof}
\end{center}
\end{figure}

\begin{figure*}
\begin{center}
\includegraphics[width=0.48\textwidth, trim=0 50 10 80, clip]{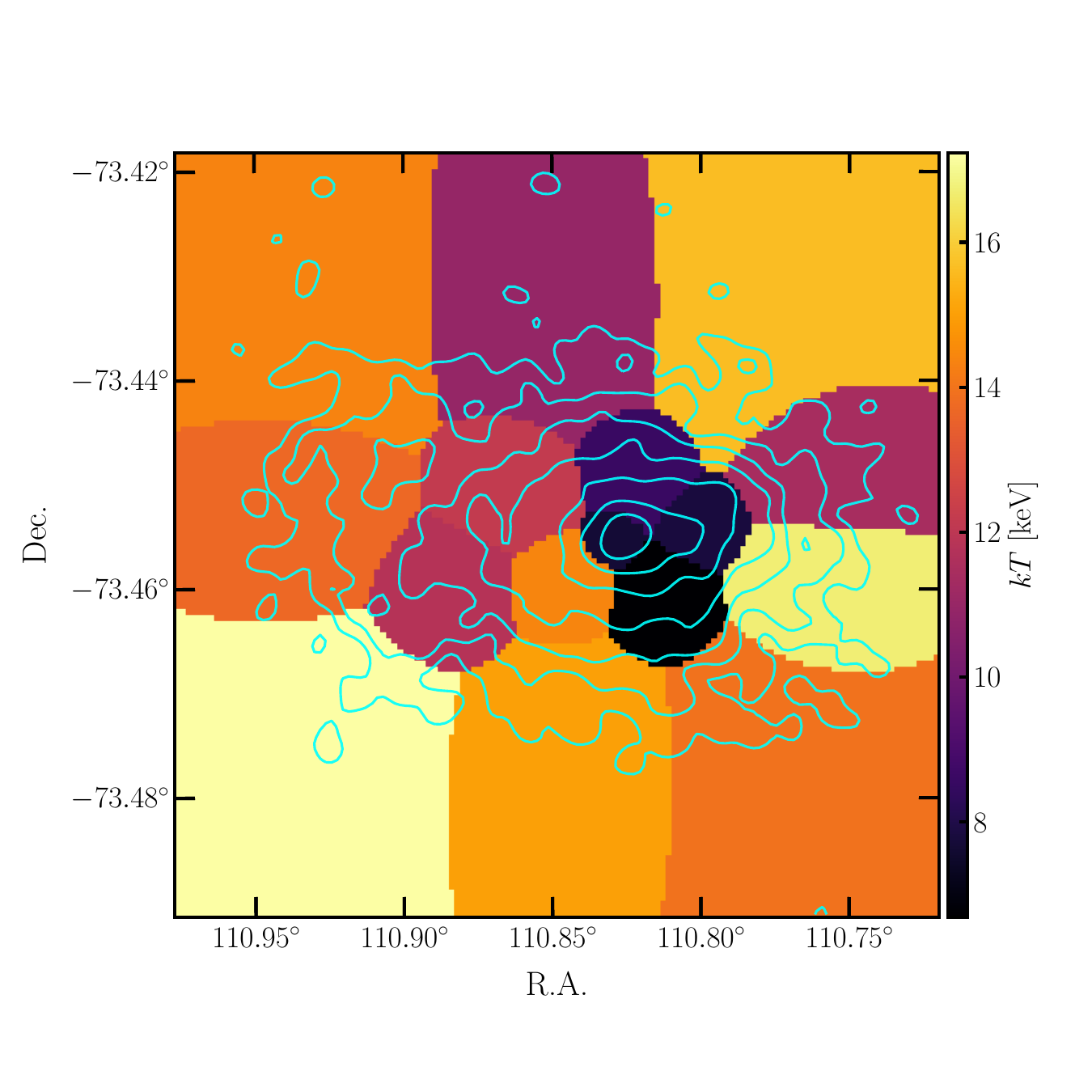}
\includegraphics[width=0.48\textwidth, trim=0 50 10 80, clip]{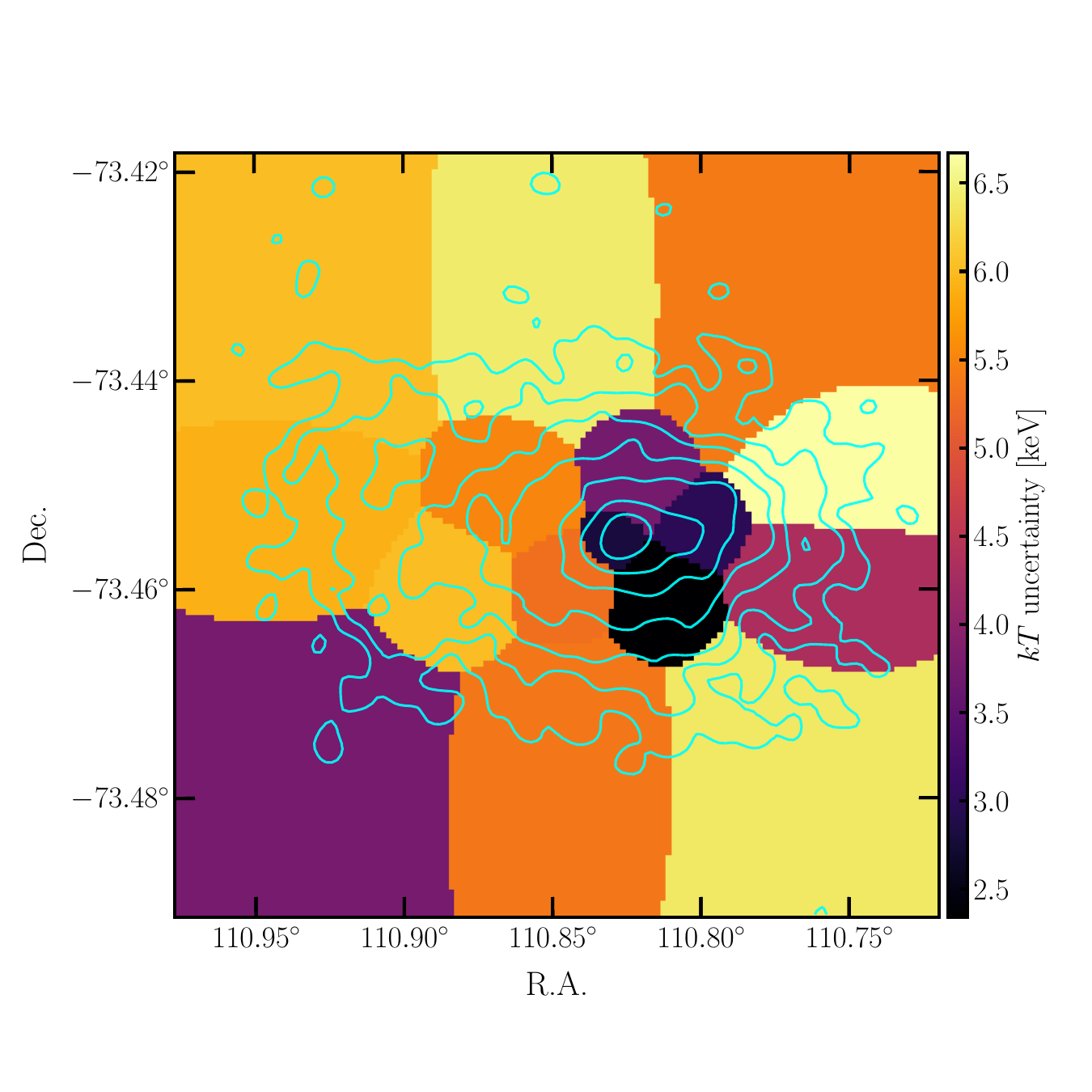}
\caption{ICM temperature distribution measured with {\sl Chandra} data. The left panel and right panel show maps of temperature and 1$\sigma$ uncertainty, respectively. The contours colored in cyan are generated from the 0.5--7~keV {\sl Chandra} image. }
\label{fig:tmap:chandra}
\end{center}
\end{figure*}

\begin{figure*}
\begin{center}
\includegraphics[width=0.49\textwidth, trim=40 45 60 30, clip]{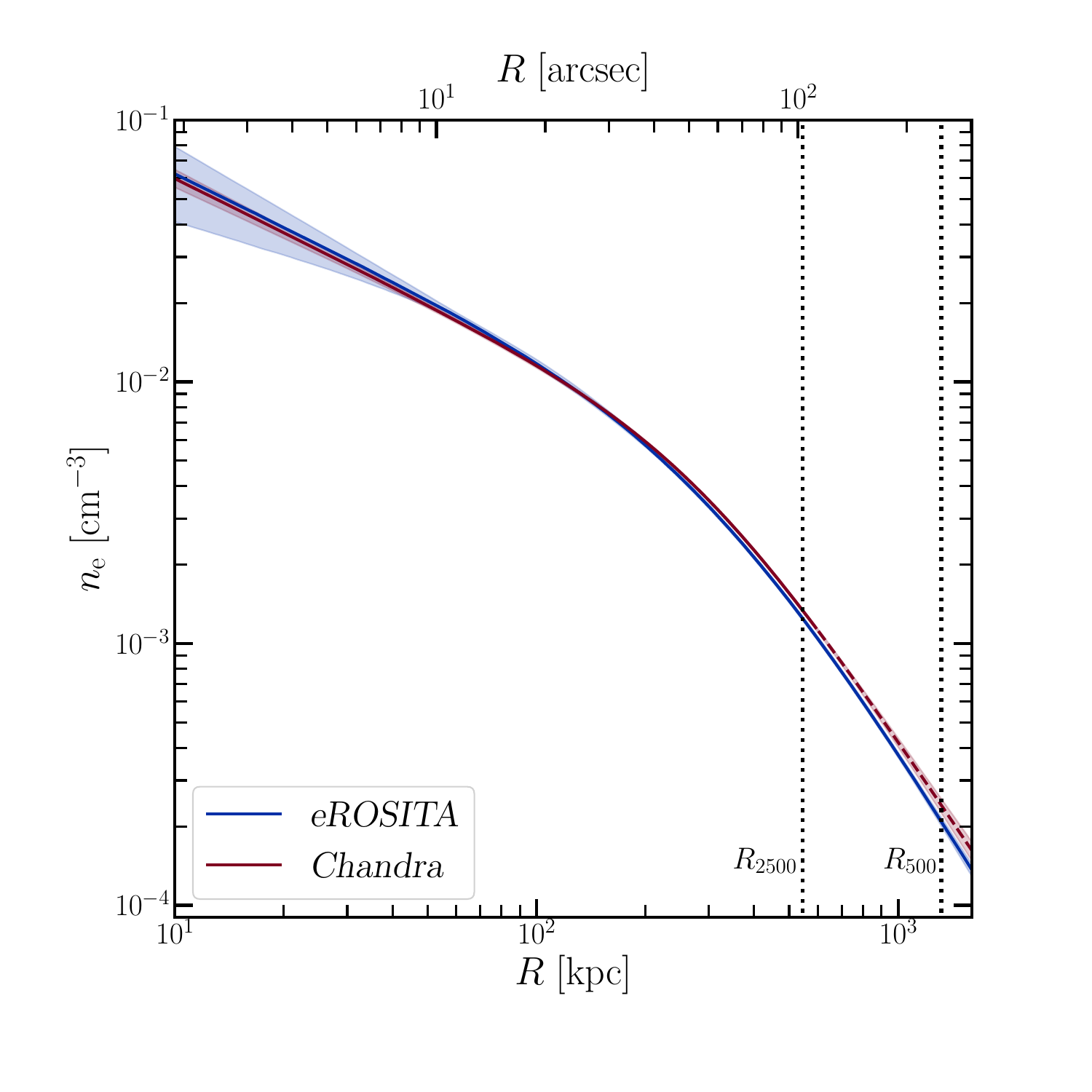}
\includegraphics[width=0.49\textwidth, trim=40 45 60 30, clip]{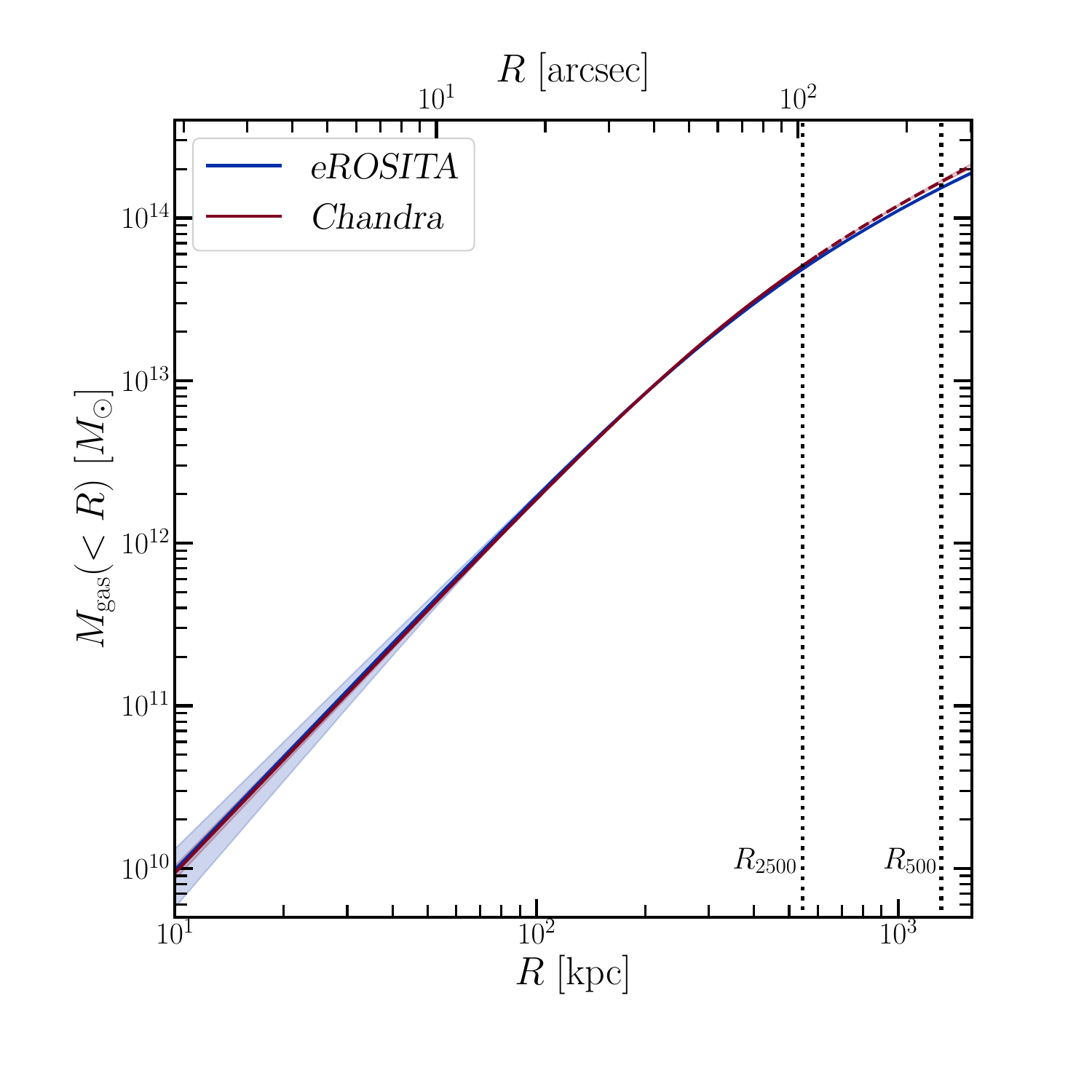}
\caption{Results of X-ray imaging analysis from {\sl eROSITA} and {\sl Chandra} using MBProj2D. {\sl Left panel:} Electron density profiles. {\sl Right panel:} Gas mass profiles. The {\sl Chandra} results above 0.6~Mpc are plotted as dashed lines.}
\label{neprof}
\end{center}
\end{figure*}

\subsection{Spectral analysis}
\label{sec:spec}
\subsubsection{eROSITA}
For the spectral analysis of {\sl eROSITA}, we choose six annuli whose outer radii are 30\arcsec, 60\arcsec, 100\arcsec, 150\arcsec, 200\arcsec, and 270\arcsec. The center of the regions is R.A. = 7:23:19.20, Dec. = $-$73:27:22.5, determined by fitting the {\sl Chandra} image (see Sect.~\ref{sec:image} and Fig.~\ref{image}). 
The spectral extraction and the computing of ancillary response files (ARFs), and redistribution matrix files (RMFs) from the seven telescope modules are performed using the {\tt eSASS} algorithm {\tt srctool}. 
The background spectrum is extracted from a concentric annulus with inner and outer radii of 3.5~Mpc and 5~Mpc, far beyond the emission of the ICM. Fitting of the background spectrum is done using the similar approach described in \citet{2022Liu}. The instrumental background due to the interactions of high-energy particles with the telescope and the satellite is obtained by fitting the filter-wheel-closed (FWC) data with a broken power-law and a series of Gaussian lines \citep{Freyberg2020}. We use the best fitting parameters obtained from the FWC data to fit the local background, but allowing a free overall normalisation. The cosmic X-ray background (CXB) is modeled by considering the contribution from the Local Hot Bubble \citep[e.g.,][]{2008Snowden,2000Kuntz}, the Galactic Halo, and unresolved point sources. The Local Hot Bubble and the Galactic Halo are modeled by an unabsorbed and absorbed {\tt apec} \citep{smith2001, 2012Foster}, and the emission due to unresolved sources is modeled by an absorbed {\tt powerlaw}, where the index is frozen to 1.46 \citep[see, e.g.,][]{2017Luo}. The Galactic hydrogen absorption is modeled using {\tt tbabs} \citep{2000Wilms}, where the hydrogen column density $n_{\rm H}$ is fixed to 22.5$\times 10^{20}$ cm$^{-2}$, according to the $n_{\rm H,tot}$ value provided by \citet{2013Willingale}. Therefore, our final CXB model is: {\tt apec+tbabs(apec+powerlaw)}. The ICM emission is modeled with a single {\tt apec} model, where the redshift is fixed at 0.39, metal abundance is fixed at 0.3. We do not make any attempt to measure the chemical property of the ICM in this work, as the metallicity of massive clusters is very well constrained from cluster center \citep[e.g.,][]{liu2018, liu2020} to outskirts \citep[e.g.,][]{mernier2018, 2017Ezer, 2016Bulbul, Simionescu2017}, and line emission is negligible in a hot cluster like J0723. 

A common issue in a combined analysis of clusters using the data from multiple telescopes is the discrepancy in the temperatures measured by different instruments. For example, \citet{2015Schellenberger} find that the cluster temperatures measured by {\sl XMM-Newton/EPIC} are significantly lower than {\sl Chandra/ACIS}. They also find that the difference is mostly due to the discrepancy in the cross-calibration in the soft band, and the temperatures measured only using the hard band are more consistent.

Therefore, in the {\sl eROSITA} analysis, we adopt the temperatures constrained from the hard band spectra (2--8~keV). Given the low number of counts we have for J0723, we are not able to perform hard-band fitting for all the bins in the temperature profile. However, this bias is systematic, and we can simply get a conversion factor: $f_T\equiv {\rm ln}T_{\rm hard}/{\rm ln}T_{\rm full}$, from a high S/N spectrum, and apply this conversion to the $T_{\rm full}$ of the spectrum with lower S/N, where $T_{\rm hard}$ cannot be constrained. To obtain the conversion factor $f_T$, we extract the spectrum within 2~arcmin to make sure the spectrum has a high S/N, and fit the full band (0.5--8~keV) and hard band (2--8~keV) spectra respectively. The best-fit temperatures are $T_{\rm hard}=9.03_{-2.60}^{+5.40}$~keV and $T_{\rm full}=4.72_{-0.32}^{+0.36}$~keV. Therefore, we have $f_T=1.417$. It should be noted that, since we assume a systematic conversion between $T_{\rm hard}$ and $T_{\rm full}$, we ignore the statistical uncertainties of $T_{\rm hard}$ and $T_{\rm full}$ in the computation of $f_T$, but only use the best-fit values. As a comparison, we also measured the temperature with {\sl Chandra} for the same region, and the result is $T_{Chandra}=9.01_{-0.56}^{+0.60}$~keV, in very good agreement with the {\sl eROSITA} hard band result. In Table~.\ref{tab:eroresu}, we list the temperature values in each bin by fitting the full band, and the modified values using the above strategy.

The modified temperature profile we obtained from the {\sl eROSITA} data is plotted in Fig.~\ref{tprof}. We note that the two innermost bins only have widths of 30$\arcsec$, thus the spectra are unavoidably affected by PSF spilling of the adjacent bin. However, this effect should be no more than a slight smoothing of the profile. We stress that in these regions we have {\sl Chandra} data with better resolution, thus the contribution of the {\sl eROSITA} temperature profile is more significant in the outskirts. Therefore the impact of PSF spilling can be safely ignored in these regions. 

\begin{table}
\caption{\label{tab:eroresu} Temperature profile measured by {\sl eROSITA} before and after the hard-band modification. }
\begin{center}
\begin{tabular}[width=\textwidth]{lcc}
\hline\hline
Region & $kT$ (full-band) [keV] & $kT$ (modified) [keV] \\
\hline
0$\arcsec$--30$\arcsec$ & $4.54_{-0.49}^{+0.52}$ & $8.56_{-1.27}^{+1.43}$ \\
30$\arcsec$--60$\arcsec$ & $5.54_{-0.67}^{+0.82}$ & $11.32_{-1.89}^{+2.45}$ \\
60$\arcsec$--100$\arcsec$ & $4.17_{-0.59}^{+0.80}$ & $7.57_{-1.47}^{+2.15}$ \\
100$\arcsec$--150$\arcsec$ & $4.78_{-0.93}^{+1.23}$ & $9.21_{-2.43}^{+3.52}$ \\
150$\arcsec$--200$\arcsec$ & $4.11_{-1.03}^{+2.00}$ & $7.43_{-2.48}^{+5.60}$ \\
200$\arcsec$--270$\arcsec$ & $3.75_{-1.09}^{+2.23}$ & $6.53_{-2.52}^{+6.11}$ \\
\hline
\end{tabular}
% \tablefoot{ }
\end{center}
\end{table}
\begin{figure*}
\begin{center}
\includegraphics[width=0.49\textwidth, trim=40 45 60 30, clip]{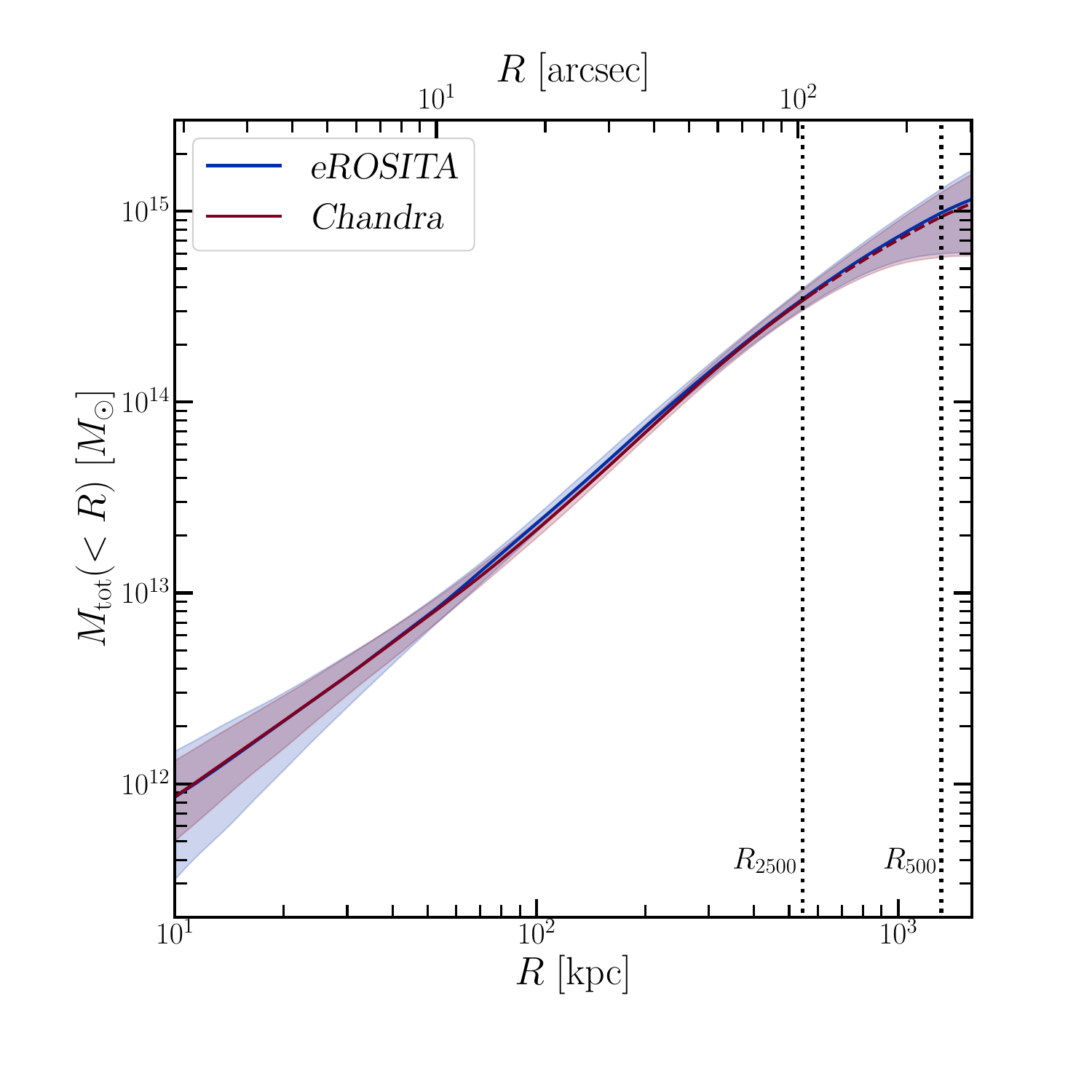}
\includegraphics[width=0.49\textwidth, trim=40 45 60 30, clip]{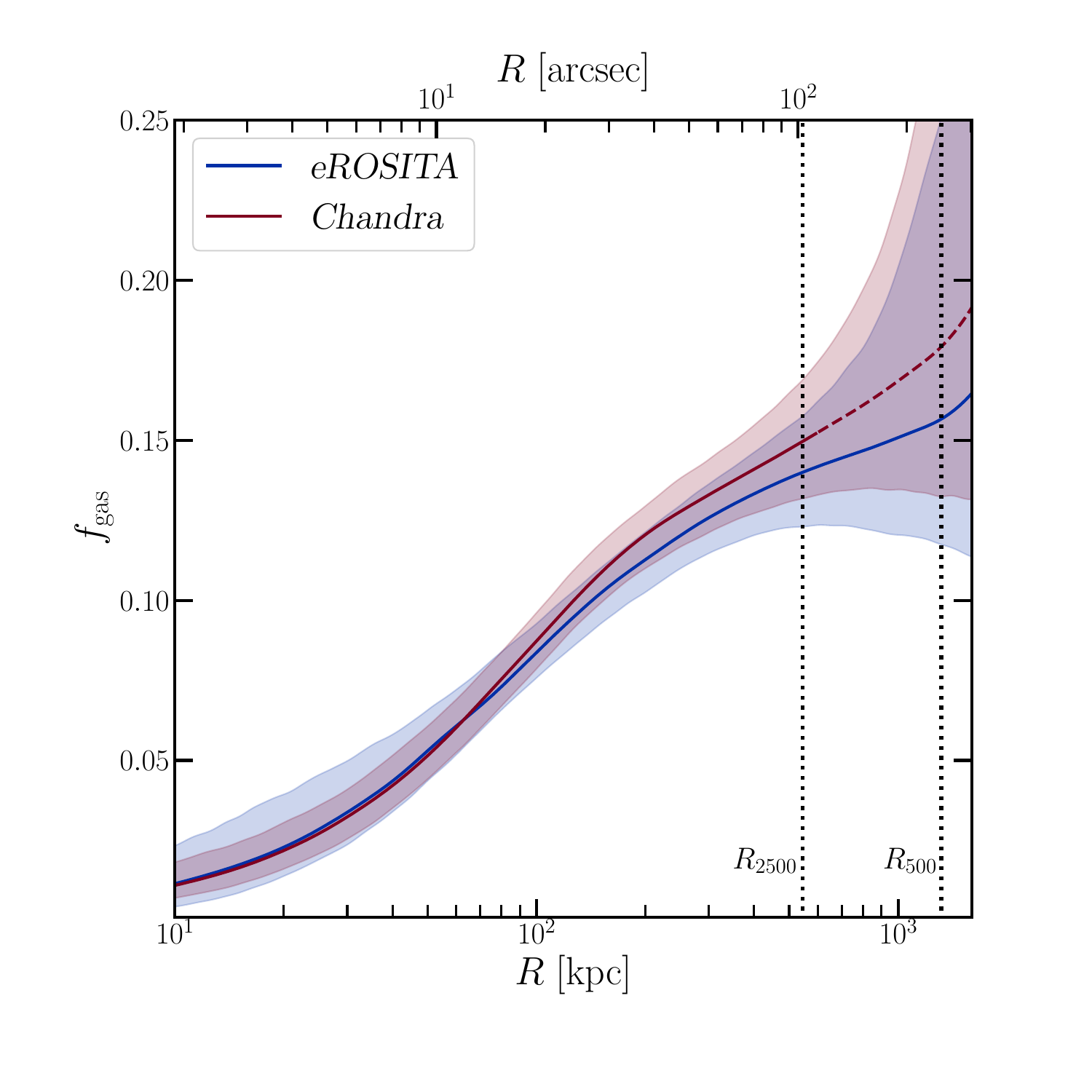}
\caption{Results of hydrostatic analysis. {\sl Left panel:} hydrostatic mass profiles of J0723 using the best-fit model of the combined temperature profile (Fig.~\ref{tprof}) and the density profile of {\sl eROSITA} (blue) and {\sl Chandra} (red). {\sl Right panel:} gas mass fraction ($f_{\rm gas}$) profiles. The two dotted lines indicate the positions of $R_{500}$=1.32~Mpc and $R_{2500}$=0.54~Mpc, determined from the hydrostatic mass profile measured by {\sl eROSITA} (blue line in the left panel). }
\label{hydromass}
\end{center}
\end{figure*}

\subsubsection{Chandra}
For the spectral analysis of {\sl Chandra} data, we increased the spatial resolution of the temperature profile in the central region ($R<50\arcsec$) with respect to the {\sl eROSITA} result. The ARF and RMF of each spectrum were computed with
the commands {\tt mkarf} and {\tt mkacisrmf}. The background spectra were extracted from the
`blank sky' files, and processed using the {\tt blanksky} script (we used the default options with {\tt weight\_method} `particle' and {\tt bkgparams = [energy = 9000:12000]}). We used the full energy range (0.5--7~keV) for the fit. We also did the same analysis by using only the hard band (2--7~keV) as we did for {\sl eROSITA}. However, we find no significant difference between the $T_{\rm hard}$ and $T_{\rm full}$ for {\sl Chandra}. This is probably because of the much smaller effective area of {\sl Chandra} in the soft band, which does not contribute much to the fit. We, therefore, adopted the full band fitting results for {\sl Chandra}. The temperature profile from {\sl Chandra} data is plotted in Fig.~\ref{tprof}. 

Using {\sl Chandra} data, we also perform a spatially-resolved spectral analysis to measure the 2D distribution of temperature, to search for any peculiar structure and inspect the consistentcy with the 1D profile. The regions for the 2D analysis are selected using the Voronoi tessellations method \citep{2003Cappellari}. Each region contains $\sim$300 net counts in the energy range 0.5--7~keV. The temperature map and its uncertainty are shown in Fig.~\ref{fig:tmap:chandra}. We find that the 1D temperature file is in broad consistent with the 2D map. A clear cool core at $\sim7$~keV can be observed in the temperature map, in accordance with the temperature profile. 

\subsubsection{Combined temperature profile}
In general, the results of {\sl eROSITA} and {\sl Chandra} agree well with each other, particularly in the central regions. In the outskirts, both datasets have large error bars. J0723 hosts a weak cool core, with $T_{\rm core}\sim7$~keV and $T_{\rm max}\sim$10~keV. Considering that the temperature variance within our datasets is less than a factor of 2, the difference between projected and deprojected temperature profiles is expected to be rather small, well below the statistical uncertainty. Therefore, we do not make any attempt to deproject the temperature profile.

The {\sl eROSITA} and {\sl Chandra} combined temperature profile is fit with the model presented in \citet{vikhlinin2006a}, in order to get a smooth temperature profile:
\begin{equation}
\label{eq:tprof}
T(r)=T_0 \cdot \frac{(r/r_{\rm cool})^{a_{\rm cool}}+T_{\rm min}/T_0}{(r/r_{\rm cool})^{a_{\rm cool}}+1} \cdot \frac{(r/r_t)^{-a}}{[1+(r/r_t)^b]^{c/b}}.
\end{equation}
The fitting is done using the MCMC tool of \citet{Foreman2013}. The best-fit profile is shown in Fig.~\ref{tprof}. The best-fit parameters of the model is listed in Table~\ref{para}.

\begin{table*}
\caption{\label{para} Best-fit parameters and uncertainties for temperature profile. }
\begin{center}
\begin{tabular}[width=\textwidth]{cccccccc}
\hline\hline
$T_0$ & $r_{\rm cool}$ & $a_{\rm cool}$ & $T_{\rm min}$ & $r_{t}$ & $a$ & $b$ & $c$ \\
\hline
[keV] & [Mpc] & [-] & [keV] & [Mpc] & [-] & [-] & [-] \\
\hline
$14.3_{-4.2}^{+12.5}$ & $0.11_{-0.07}^{+0.06}$ & $0.38_{-0.28}^{+0.38}$ & $5.7_{-3.1}^{+3.0}$ & $2.8_{-1.4}^{+1.5}$ & $-0.07_{-0.12}^{+0.09}$ & $2.5_{-1.4}^{+1.7}$ & $4.3_{-3.0}^{+3.8}$ \\
\hline
\end{tabular}
% \tablefoot{ }
\end{center}
\end{table*}

\subsection{Image fitting with MBProj2D}
\label{sec:image}
To measure the gas density profile, we perform imaging analysis using the MultiBand Projector 2D (MBProj2D) tool\footnote{https://github.com/jeremysanders/mbproj2d}. MBProj2D \citep{2018Sanders} is a code which forward-models background-included X-ray images of galaxy clusters to fit cluster and background emission simultaneously and measures the profiles of ICM properties including density, flux, luminosity, etc. By using a single band image, MBProj2D is only sensitive to the density of the ICM. By using multiple bands from soft to hard and given enough counts, MBProj2D is also able to model the temperature variation within the cluster. When sufficient energy bands are provided, MBProj2D can give equivalent results as expected from spatially-resolved spectral analysis, e.g., metallicity and temperature profiles of the ICM. By assuming hydrostatic equilibrium, MBProj2D can also measure hydrostatic mass profile. We refer the readers to \citet{2018Sanders} for more details about MBProj2D \footnote{https://mbproj2d.readthedocs.io/en/latest/}.

In this work, since we already have the temperature profile from spectral analysis, we only use the basic function of MBProj2D, to measure the density profile using multiple band images, without assuming a hydrostatic mass model, and then derive the model-independent hydrostatic mass profile backwardly using the temperature and density profiles. For {\sl eROSITA} data, we create images and exposure maps in the following energy bands (in units of keV): [0.3--0.6], [0.6--1.0], [1.0--1.6], [1.6--2.2], [2.2--3.5], [3.5--5.0], [5.0--7.0], using the {\tt evtool} command in {\tt eSASS}. The size of the image is 6~Mpc $\times$ 6~Mpc, which is large enough to include the background. Point sources within the image are masked following the same approach in the spectral analysis. PSF and ARF variations across different bands are considered properly. The model of density profile is from \citet{vikhlinin2006a}, but without the second $\beta$ component:
\begin{equation}
n_{\mathrm p}n_{\mathrm e} = n_0^2 \cdot \frac{(r/r_c)^{-\alpha}}{(1+r^2/r_c^2)^{3\beta-\alpha/2}}\frac{1}{(1+r^{\gamma}/{r_s}^{\gamma})^{\epsilon/\gamma}},
\end{equation}
where $n_{\rm e}$ and $n_{\rm p}$ are electron density and proton density, and we assume $n_{\rm e}=1.21n_{\rm p}$. $\gamma$ is fixed at 3. $n_0$, $r_c$, $\alpha$, $\beta$, $r_s$ and $\epsilon$ are free parameters. 

A similar analysis was done also for the {\sl Chandra} data. The following energy bands (in units of keV) are used for {\sl Chandra} analysis: [0.5--0.75], [0.75--1.0], [1.0--1.25], [1.25--1.5], [1.5--2.0], [2.0--3.0], [3.0--4.0], [4.0--5.0], [5.0--6.0], [6.0--7.0]. Limited by the FOV of {\sl Chandra}, the size of the image is reduced to 4~Mpc $\times$ 4~Mpc. However, this is still large enough to contain the background emission in the fitting. The {\sl Chandra} exposure map computed by {\tt CIAO} is always folded by the effective area. In order to get the unfolded exposure map as required by MBProj2D, we extracted a spectrum in the center of the image and renormalised the folded exposure maps using the exposure time information of the spectrum. 

The electron density profiles measured using MBProj2D are plotted in Fig.~\ref{neprof}. The best-fit parameters are provided in Table~\ref{para_ne}. Also shown in Fig.~\ref{neprof} are the cumulative gas mass profiles. The results from {\sl eROSITA} and {\sl Chandra} agree very well. There is only a very small discrepancy in the density profile in the outskirts, which is acceptable considering that the {\sl Chandra} data is too shallow in these regions, thus the {\sl Chandra} result is probably an extrapolation of the best-fit model dominated by the central region. 

\begin{table*}
\caption{\label{para_ne} Best-fit parameters and uncertainties for electron density profiles. }
\begin{center}
\begin{tabular}[width=\textwidth]{ccccccc}
\hline\hline
Instrument & $n_0$ & $r_c$ & $\alpha$ & $\beta$ & $r_s$ & $\epsilon$ \\
\hline
 & [$10^{-3} {\rm cm}^{-3}$] & [kpc] & [-] & [-] & [kpc] & [-] \\
\hline
{\sl eROSITA} & $7.6_{-3.0}^{+5.9}$ & $210.0_{-81.0}^{+119.8}$ & $1.42_{-0.49}^{+0.28}$ & $0.55_{-0.16}^{+0.13}$ & $519.4_{-186.4}^{+316.4}$ & $1.32_{-0.89}^{+0.75}$ \\
{\sl Chandra} & $5.6_{-1.6}^{+1.6}$ & $306.4_{-69.7}^{+128.3}$ & $1.39_{-0.10}^{+0.09}$ & $0.60_{-0.15}^{+0.16}$ & $274.6_{-65.3}^{+220.1}$ & $0.64_{-0.92}^{+0.87}$ \\
\hline
\end{tabular}
% \tablefoot{ }
\end{center}
\end{table*}

\begin{table*}
\caption{\label{tab:resu} Comparison between {\sl eROSITA} and {\sl Chandra} in the measurements of cluster masses within $R_{500}$ and $R_{2500}$. }
\label{resu}
\begin{center}
\begin{tabular}[width=\textwidth]{lcccccc}
\hline\hline
Instrument & $R_{500}$ & $M_{\rm gas,500}$ & $M_{500}$ &  $R_{2500}$ & $M_{\rm gas,2500}$ & $M_{2500}$  \\
\hline
 & [Mpc] & [10$^{14}~M_{\odot}$] & [10$^{14}~M_{\odot}$] & [Mpc] &  [10$^{14}~M_{\odot}$] & [10$^{14}~M_{\odot}$]  \\
\hline
{\sl eROSITA} & $1.32\pm0.23$ & $1.53\pm0.30$ & $9.79\pm5.07$ & $0.54\pm0.04$ & $0.49\pm0.06$ & $3.47\pm0.79$ \\
{\sl Chandra} & $1.29\pm0.22$ & $1.63\pm0.33$ & $9.15\pm4.65$ & $0.54\pm0.04$ & $0.50\pm0.06$ & $3.36\pm0.74$ \\
\hline
\end{tabular}
% \tablefoot{ }
\end{center}
\end{table*}
\subsection{Hydrostatic mass}
\label{sec:hydro}

\begin{figure*}
\begin{center}
\includegraphics[width=0.49\textwidth, trim=40 45 60 30, clip]{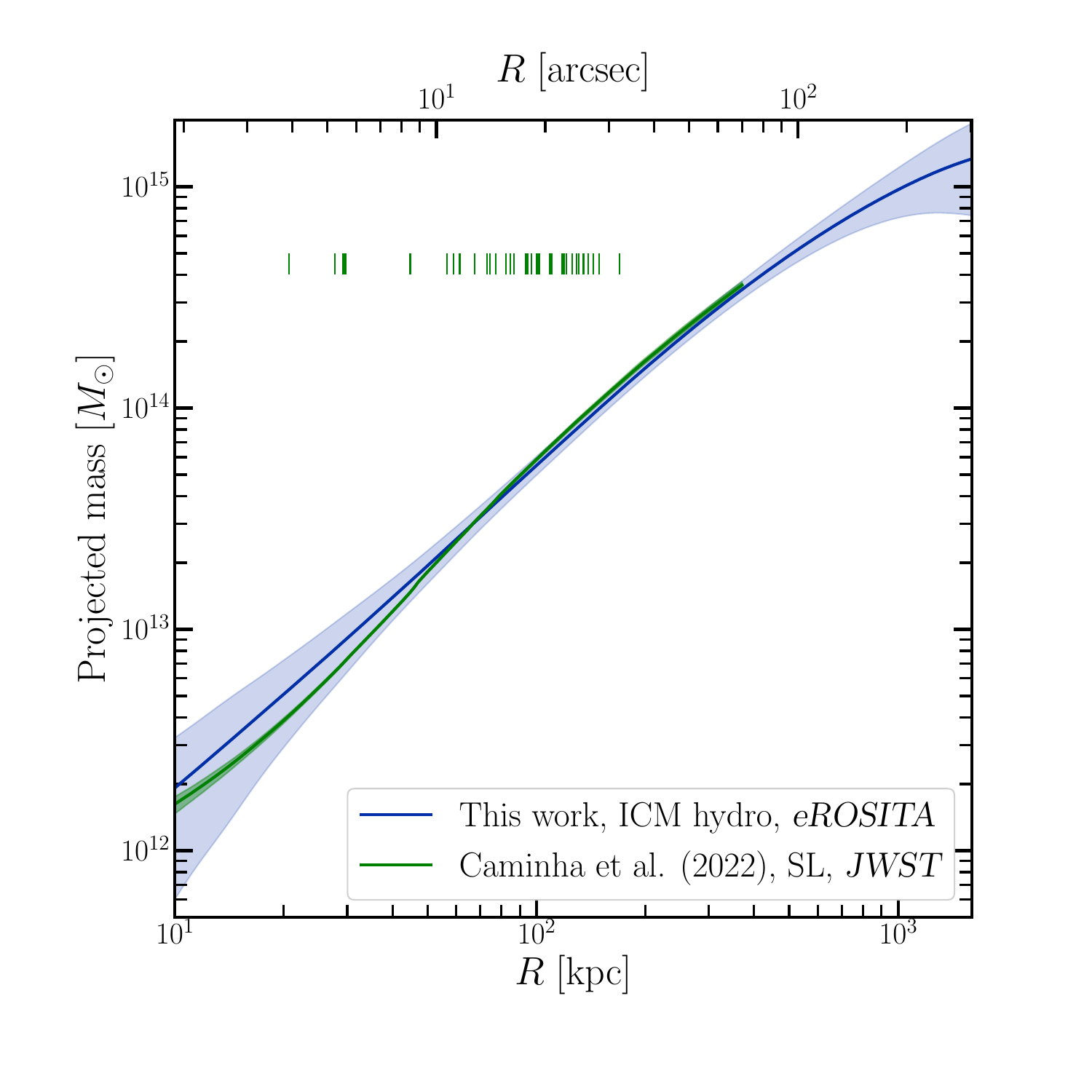}
\includegraphics[width=0.49\textwidth, trim=40 45 60 30, clip]{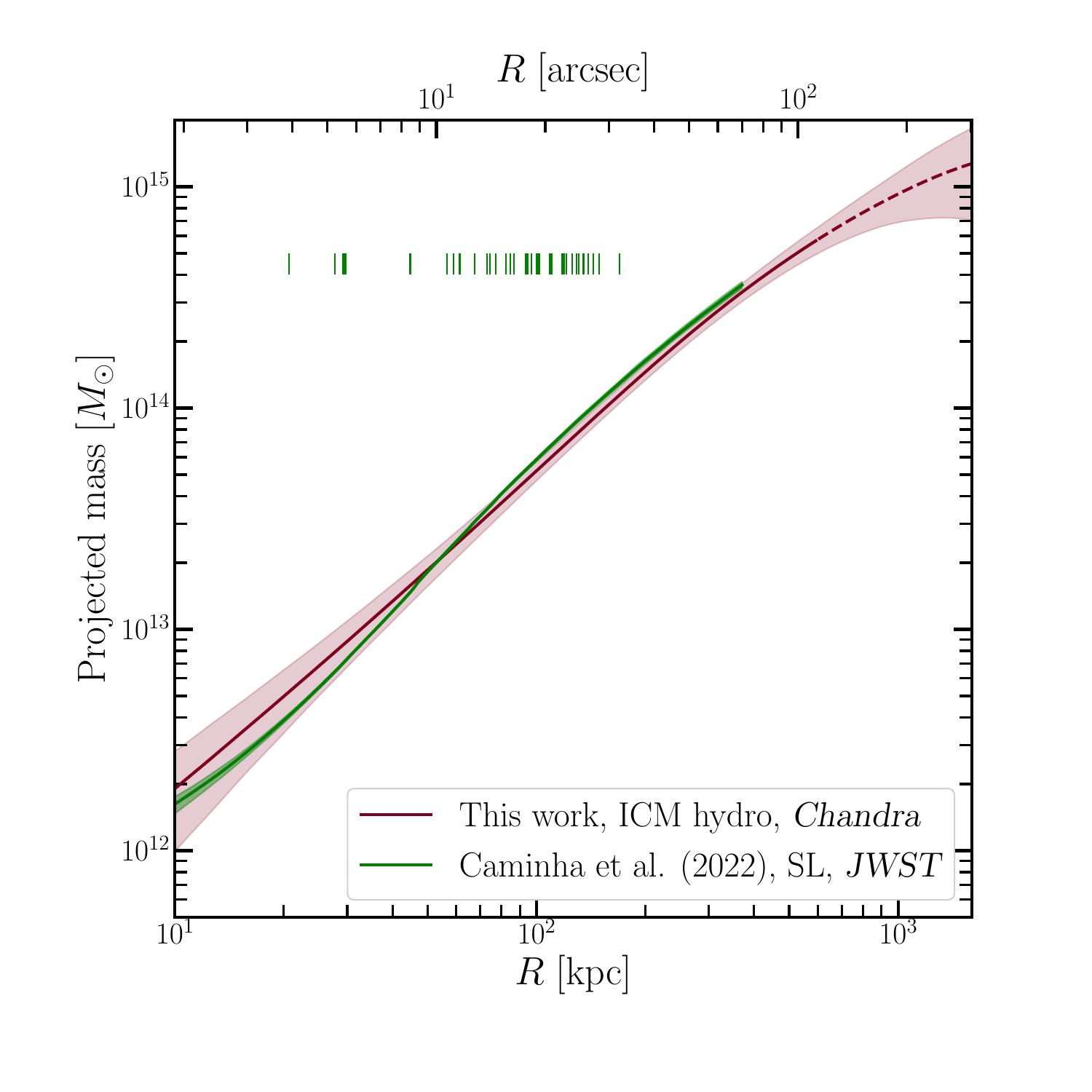}
\caption{Line-of-sight projected mass profiles of J0723 measured by {\sl eROSITA} ({\sl left panel}) and {\sl Chandra} ({\sl right panel}). The green curve shows the mass distribution from strong lensing analysis with multiple images discovered by {\sl JWST} \citep[see Fig.A.1 of][]{2022Caminha}. The green bars shows the positions of the multiple images used in the strong lensing analysis. }
\label{projmass}
\end{center}
\end{figure*}

\begin{figure*}
\begin{center}
\includegraphics[width=0.49\textwidth, trim=40 45 60 30, clip]{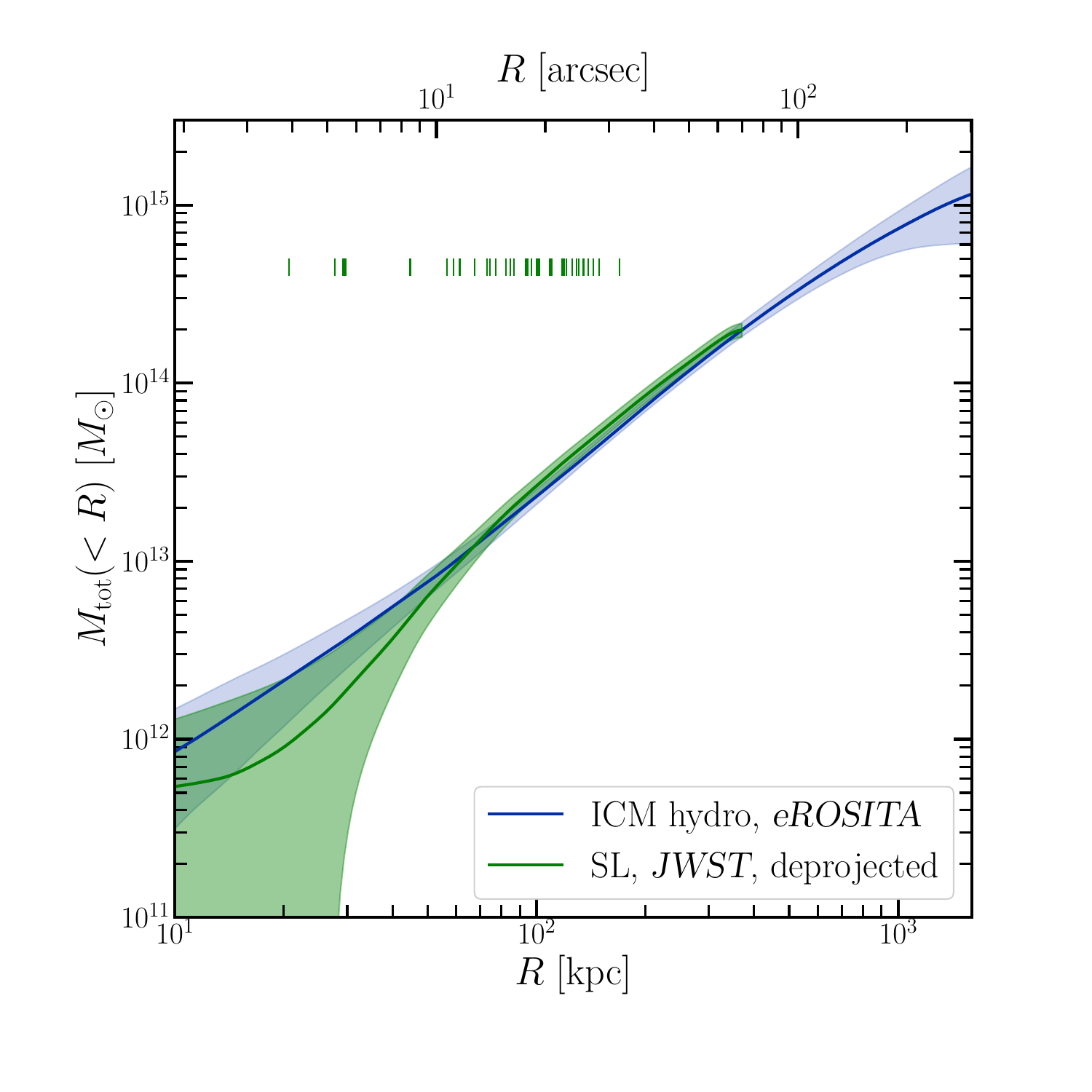}
\includegraphics[width=0.49\textwidth, trim=40 45 60 30, clip]{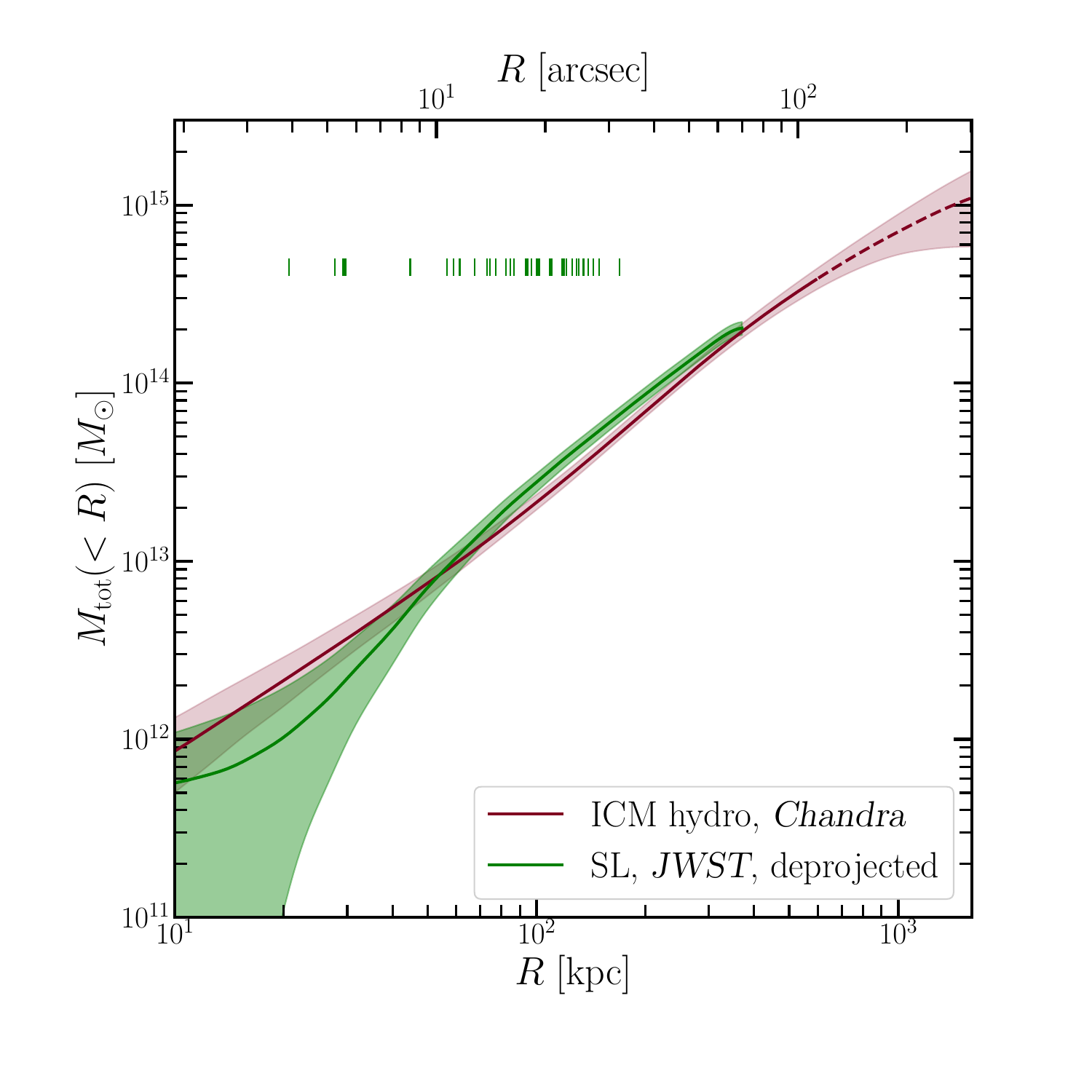}
\caption{Comparison between the hydrostatic (hydro) mass profile (blue and red curves) and the strong lensing (SL) mass profile in \citet{2022Caminha} after deprojection (green curves). Left panel is for {\sl eROSITA} and right panel is for {\sl Chandra}. }
\label{hydromass_lens}
\end{center}
\end{figure*}

With the ICM temperature profile and density profile, we derive the hydrostatic mass profile using the hydrostatic equilibrium equation:
\begin{equation}
M(<r)=-\frac{k_{\mathrm{B}} T r}{G \mu m_{\mathrm{p}}}\left(\frac{\mathrm{d} \ln \rho_{\mathrm{g}}}{\mathrm{d} \ln r}+\frac{\mathrm{d} \ln T}{\mathrm{~d} \ln r}\right),
\end{equation}
where $k_{\mathrm B}$ is the Boltzmann constant, $G$ is the gravitational constant, $\mu=0.6$ is the mean atom weight, $m_{\rm p}$ is proton mass. $\rho_{\mathrm g}=n_{\rm e}m_{\rm p}A/Z$ is the gas density, where $A\sim1.4$ and $Z\sim1.2$ are the average nuclear charge and mass for ICM with 0.3 solar abundance. The uncertainty on the mass profile was computed by randomly picking 2000 samples in the Markov chains of the temperature and density profiles.

The hydrostatic mass profiles measured by {\sl eROSITA} and {\sl Chandra} are plotted in the left panel of Fig.~\ref{hydromass}. We note that, since we are using the combined temperature profile, the difference in the two profiles results only from the density profiles. We find that the results obtained from {\sl eROSITA} and {\sl Chandra} are in very good agreement, despite the small discrepancies in the very center ($<50$~kpc) and the outskirts ($>1$~Mpc), which are well within the error bar. From the mass profile measured by {\sl eROSITA} (see the blue line in Fig.~\ref{hydromass}), we obtained $R_{2500}=0.54\pm0.04$~Mpc, $R_{500}=1.32\pm0.23$~Mpc, and the masses within the two radii $M_{2500}=(3.5\pm0.8)\times 10^{14}~M_{\odot}$ and $M_{500}=(9.8\pm5.1)\times 10^{14}~M_{\odot}$. The values computed from the {\sl Chandra} results are fully consistent (see Table.~\ref{tab:resu}). The error bars in $R_{500}$ and $M_{500}$ are large, due to the relatively larger uncertainty of the temperature in the outskirts. We also find a perfect agreement with the published results of \citet{2020Lovisari}, who measured $M_{500}=10.1_{-1.2}^{+1.6}\times 10^{14} M_{\odot}$ using the archival data of {\sl XMM-Newton}. The temperature measured by \citet{2020Lovisari}, $kT_{500}=7.53_{-0.53}^{+0.53}$~keV, is slightly lower compared to our temperature profile, but still consistent within $1\sigma$. We also compare the mass of J0723 we measured in this work to the published $L-M$ scaling relation of galaxy clusters. From the luminosity profile measured with MBProj2D using {\sl eROSITA} data, we obtain $L_{500}=11.2_{-0.1}^{+0.2}\times 10^{44}$erg/s in the 0.5--2~keV band. Adopting the $L-M$ scaling relation for high redshift and massive clusters in \citet{bulbul2019}, the corresponding mass is $M_{\rm 500, SR}=10.7_{-0.1}^{+0.1}\times 10^{14}M_{\odot}$, in very good agreement with our results.

With the hydrostatic mass profile and gas mass profile, we also computed the profile of gas mass fraction ($f_{\rm gas}$), which is shown in the right panel of Fig.~\ref{hydromass}. With {\sl eROSITA} data, we measure $f_{\rm gas,500}=0.157_{-0.040}^{+0.103}$ and $f_{\rm gas,2500}=0.140_{-0.016}^{+0.017}$. From the profile, we observe a clear trend that $f_{\rm gas}$ decreases from $>10\%$ in the outskirts ($>1$~Mpc) to $<5\%$ in the core ($<50$~kpc), despite that the former has large uncertainty. This is consistent with the picture that in massive galaxy clusters, the dark matter halo and stellar mass are more concentrated than the hot gas, thus dominating the total mass in the center, while the mass budget tends to be more consistent with the average of the Universe when extending to larger radii \citep[see, e.g.,][]{2013Eckert, 2016Planck}. 

In order to compare with the strong lensing results, we compute the line-of-sight projected mass using our hydrostatic mass profiles. The projected mass profiles are shown in Fig.~\ref{projmass}. We compare our results with the recent strong lensing results published in \citet{2022Caminha}, who measure the cluster mass distribution with 46 multiple images from 16 background sources. We find remarkable consistency between the {\sl eROSITA} hydrostatic mass and the strong lensing mass at all radii. At the estimated Einstein radius 128~kpc \citep{2022Caminha}, {\sl eROSITA} measured a projected hydrostatic mass of $(8.0\pm0.7)\times 10^{13} M_{\odot}$, in good agreement with the result reported in \citet{2022Caminha}: $(8.6\pm0.2)\times 10^{13} M_{\odot}$. The measurement of {\sl Chandra} at the same radius: $(7.6\pm0.7)\times 10^{13} M_{\odot}$, is only slightly ($\sim 1\sigma$) lower than the strong lensing result, while they tend to be more consistent at larger radii. Using the same {\sl JWST} data, another team measured a strong lensing mass of $(7.6\pm0.2)\times 10^{13} M_{\odot}$ at 128~kpc \citep{2022Mahler}, in tension with the result of \citet{2022Caminha} at $\sim4\sigma$. On the other hand, we note that the result of \citet{2022Mahler} is consistent with our results for both {\sl eROSITA} and {\sl Chandra} within $1\sigma$. 

As a further check, we also deproject the strong lensing mass profile from \citet{2022Caminha} using the hydrostatic mass profiles measured by {\sl eROSITA} and {\sl Chandra} respectively, and compare it with the hydrostatic mass profiles. From the comparison shown in Fig.~\ref{hydromass_lens}, the consistency between {\sl eROSITA} and strong lensing results still remains good. For {\sl Chandra}, the two profiles are also consistent within $\sim 1\sigma$.

We also remark that, there are many other strong lensing analyzes published in literature, both before and after {\sl JWST}, thanks to the rapid development of lensing modelization techniques and optical spectroscopic surveys in recent years \citep[e.g.,][]{2022Fox, 2022Golubchik, 2022Sharon}. Despite that most of these works provide consistent results in the center ($<200$~kpc), there are tensions at larger radii due to the different tools and data they used. However, here we refrain from comparing our results with all the available strong lensing results, and from discussing the tensions between them, which are beyond the goal of this work.

\section{Discussion}
\label{sec:discussion}

In this section, we discuss the systematics and caveats in our analysis. A possible source of bias to our results is the deviation from hydrostatic equilibrium of the ICM in J0723. This so-called ``hydrostatic mass bias'' can be caused by major/minor mergers \citep[e.g.,][]{markevitch2002, vikhlinin2001}, core sloshing \citep[e.g.,][]{markevitch2001, sanders2005}, gas turbulence \citep[only a few percent, see][]{2019Eckert}, possible global rotation \citep{Liu2019b}, and other non-gravitational processes such as AGN feedback \citep{fabian2012}. On the {\sl Chandra} image, there are no signs of violent mergers. Weak emission excess is visible $\sim400$~kpc to the northeast of the X-ray peak (see the right panel of Fig.~\ref{image}), possibly associated with a gas clump, but it is too close to the cluster center and too faint, thus only has negligible impact on the hydrostatic analysis. The offset between the X-ray peak and the position of the brightest cluster galaxy (BCG) is $\sim6\arcsec$, corresponding to $\sim30$~kpc, or $\sim0.02R_{500}$, implying that J0723 is in a relaxed dynamical status \citep[see, e.g.,][]{2016Rossetti,2021Seppi}. Moreover, with a visual inspection on the {\sl Chandra} image, we do not find any significant structure related with AGN mechanical feedback, such as cavities. Although such a visual inspection is limited by the number of photons, and it is hard to quantify the existence/non-existence of cavities, we can conservatively conclude that the impact of feedback is not strong enough to affect our results. 

On the {\sl Chandra} image there is also a mild signal of asymmetry in the central region, which may indicate the sloshing of the core. On the temperature map (see Fig.~\ref{fig:tmap:chandra}), the coolest region seems to be closer to the west edge of the core, instead of the X-ray peak, which indicates a possible sloshing of the cool core, consistent with the surface brightness image. However, limited by the data quality, we are not able to make further analysis on the dynamical status of the core. In any case, we can confirm that J0723 is well virialised, and the impact of non-hydrostatic gas on the measurement of the hydrostatic mass can be ignored. Most studies so far suggest a hydrostatic mass bias of a few to $\sim20$ percent, from the most relaxed clusters to the extremely disturbed ones \citep[see][for a review]{2019Pratt}. The bias is sensitive to the dynamic status of the cluster, and the radius within which the masses are measured. For J0723, the hydrostatic masses we measured are within $1\sigma$ with the strong lensing masses which are only available in the central region. In the outskirts ($\sim R_{500}$), the uncertainty in the hydrostatic mass profile increases to $>50\%$, far beyond the typical value of hydrostatic bias for relaxed clusters, and the strong lensing measurements are not available. Therefore, with the results in this work, we do not observe a statistically significant hydrostatic bias. 

\begin{figure}
\begin{center}
\includegraphics[width=0.49\textwidth, trim=30 45 45 60, clip]{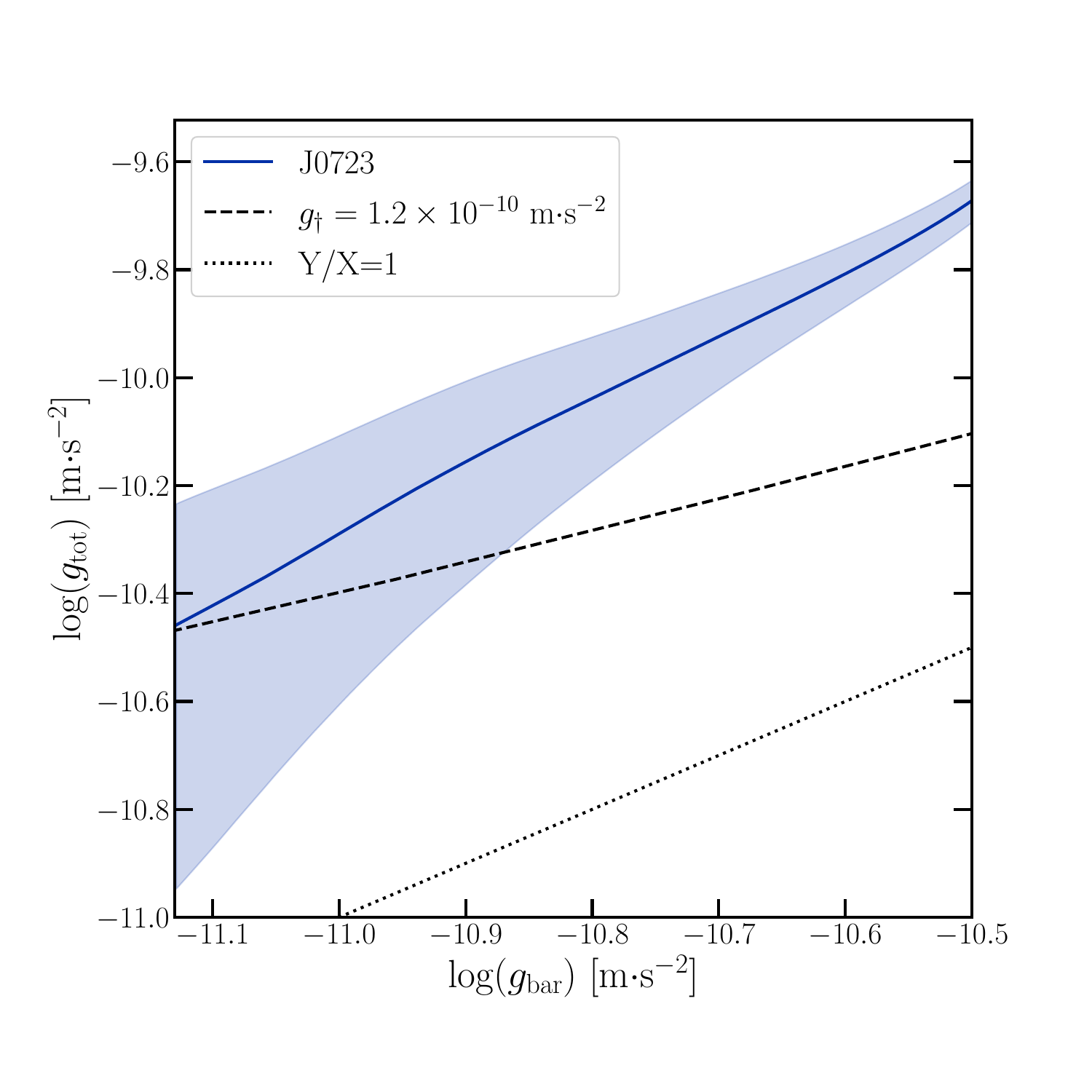}
\caption{Radial acceleration relation of J0723. The dashed line shows the RAR in spiral galaxies obtained by \citet{2016McGaugh}. The dotted line indicates the $Y=X$ relation. }
\label{rar}
\end{center}
\end{figure}

With the gas mass and hydrostatic mass profiles, we also make the attempt to study the radial acceleration relation (RAR) in J0723. RAR describes the relation between the centripetal accelerations due to the total mass ($g_{\rm tot}=GM_{\rm tot}/r^{2}$) and baryonic mass ($g_{\rm bar}=GM_{\rm bar}/r^{2}=G(M_{\rm gas}+M_{\ast})/r^{2}$) without introducing dark matter. Studying RAR in gravitationally bound systems is useful to verify modified gravity theories, such as the Modified Newtonian Dynamics \citep[MOND,][]{1983Milgrom}. A universal RAR across all mass scales is regarded as an important support for MOND. So far, a tight and universal RAR has been found in a number of studies on galaxies \citep{2016McGaugh,2017Lelli,2018Li,2021Brouwer}. For example, \citet{2016McGaugh} found that the RAR of 153 spiral galaxies fits well with the following model: 
\begin{equation}
\label{eq:rar}
\frac{g_{\rm tot}}{g_{\rm bar}}=\frac{1}{1-e^{-\sqrt{g_{\rm bar}/g_{\dagger}}}}, 
\end{equation}
where $g_{\dagger}=1.20\pm0.02$ (random) $\pm$ 0.24 (systematic) $\times 10^{-10} {\rm m \cdot s}^{-2}$.

On the other hand, the RAR in galaxy clusters is less clear. Not only is the normalization of acceleration higher than what is found for galaxies \citep[e.g.,][]{2020Tian}, but also the overall shape of the observed cluster RAR clearly deviates from the model in Eq.~\ref{eq:rar}, implying a completely different RAR between clusters and galaxies \citep[e.g.,][]{2022Eckert,2022Tam}.

To compute the RAR in J0723, we use the hydrostatic mass profile and gas mass profile obtained from {\sl eROSITA}. Baryonic mass in the central region is dominated by stars in the BCG and intracluster light (ICL), where we do not have constraint on the stellar mass profile. We thus ignore the core with a size of 200~kpc in the following discussion. Beyond the core, the contribution of stellar mass in the total mass budget is roughly a constant at $\sim2\%$ up to 2.5~Mpc \citep[see, e.g.,][]{2015Andreon,2022Eckert}. Therefore, we approximate the stellar mass profile $M_{\ast}(r)$ as $0.02\times M_{\rm tot}(r)$, adding on a systematic uncertainty of 50\%. Since stellar mass only contributes 10\%--15\% in baryonic mass at large radii \citep{2022Eckert}, this rough estimation will not significantly affect our results.

The RAR in J0723 is shown in Fig.~\ref{rar}. Depite the large errorbar and the narrow range in $g_{\rm bar}$ and $g_{\rm tot}$, we clearly observe that the J0723 RAR is inconsistent with the model of \citet{2016McGaugh} for spiral galaxies. The discrepancy is so large that it cannot be explained by either the possible biases or systematics in our analysis, such as the uncertainty in stellar mass and hydrostatic bias, or the 0.12 dex scatter in the \citet{2016McGaugh} relation. This is in broad agreement with the results for X-COP clusters \citep{2022Eckert} and BAHAMAS simulation \citep{2022Tam}. Due to the removal of the core, we have no constraint on $g_{\rm bar}$ at high acceleration, where \citet{2022Eckert} observed that the RAR of X-COP clusters eventually catched up with the \citet{2016McGaugh} relation for galaxies. However, this trend is largely predictable as the core region with high acceleration is dominated by the BCG, while ICM only constitutes a small fraction in the total baryonic mass. In summary, we find that the RAR found in galaxies cannot fit the data of the cluster J0723, thus our results do not support a universal RAR across all mass scales.  

\section{Conclusions}
\label{sec:conclusions}
In this work, we perform X-ray analysis on the galaxy cluster SMACS~J0723.3-7327, using the all-sky survey data from the {\sl SRG/eROSITA} telescope and {\sl Chandra} observatory. The azimuthally-average profiles of ICM temperature and density are measured with high accuracy from cluster center to the outskirts ($\sim R_{500}$), from which the hydrostatic mass profile is derived. While the high-angular-resolution {\sl Chandra} data has most of its contribution in measuring the temperature profile in the center, the {\sl eROSITA} data dominates the constraints from the center to the outskirts, thanks to the significantly larger effective area (the effective area ratio between {\sl eROSITA} and {\sl Chandra} in 2014 is $\sim10$ in the 0.5--2~keV band and $\sim2$ in the 2--7~keV band).
{\sl eROSITA} and {\sl Chandra} are consistent with each other in the measurements of temperature, density, gas mass, and hydrostatic mass, particularly within $\sim R_{2500}$, while the {\sl Chandra} results in the outskirts are less conclusive due to the shallow observation. The X-ray results are compared with the lens mass model constrained from strong lensing analysis with the latest {\sl JWST} data \citep{2022Caminha}. We find remarkable consistency between the X-ray hydrostatic mass profile and the strong lensing mass distribution at all radii (the latter ends at $\sim400$~kpc), both projected and deprojected. We also find that the radial acceleration relation (RAR) in J0723 is inconsistent with the RAR for spiral galaxies, implying that the latter is not a universal property of gravity across all mass scales. 

With the published results in the Early Data Release, {\sl eROSITA} has proved its capability particularly in the detection and study of low-mass clusters and groups \citep[see, e.g.,][]{2022Liu,2022Bulbul,2022Klein,2022Bahar,2022Ghirardini,2022Pasini,2022Veronica}. In this study, we have focused on J0723, a more massive cluster at high redshift ($z=0.39$) and with high temperature ($\sim10$~keV). With the eRASS:5 survey depth data, we have measured the X-ray properties of the cluster with comparable accuracy as reached by the same analysis using deeper {\sl Chandra} data. A similarly hot cluster, A3266, has been studied as a calibration target in \citet{2022Sanders}, but it is located at a much lower redshift and with a deeper pointing observation. Therefore, with this work, we further verify the capability of {\sl eROSITA}, and also the potential of the all-sky survey data, in the study of massive and distant galaxy clusters.

\begin{acknowledgement}
We thank the anonymous referee for his/her constructive comments that helped improve the paper. 
We thank G. B. Caminha for providing the JWST strong lensing model.

This work is based on data from eROSITA, the soft X-ray instrument aboard SRG, a joint Russian-German science mission supported by the Russian Space Agency (Roskosmos), in the interests of the Russian Academy of Sciences represented by its Space Research Institute (IKI), and the Deutsches Zentrum für Luft- und Raumfahrt (DLR). The SRG spacecraft was built by Lavochkin Association (NPOL) and its subcontractors, and is operated by NPOL with support from the Max Planck Institute for Extraterrestrial Physics (MPE).

The development and construction of the eROSITA X-ray instrument was led by MPE, with contributions from the Dr. Karl Remeis Observatory Bamberg \& ECAP (FAU Erlangen-Nuernberg), the University of Hamburg Observatory, the Leibniz Institute for Astrophysics Potsdam (AIP), and the Institute for Astronomy and Astrophysics of the University of Tübingen, with the support of DLR and the Max Planck Society. The Argelander Institute for Astronomy of the University of Bonn and the Ludwig Maximilians Universität Munich also participated in the science preparation for eROSITA.
\\
The eROSITA data shown here were processed using the eSASS/NRTA software system developed by the German eROSITA consortium.
\\
A.L. and E.B. acknowledge financial support from the European Research Council (ERC) Consolidator Grant under the European Union’s Horizon 2020 research and innovation programme (grant agreement CoG DarkQuest No 101002585). 

\end{acknowledgement}

\bibliography{smacsj0723}

\end{document}